\documentclass[english,aps,reprint]{revtex4-1}
\usepackage[T1]{fontenc}
\usepackage{color}
\usepackage{amsmath}
\usepackage{amssymb}
\usepackage{graphicx}
\PassOptionsToPackage{normalem}{ulem}
\usepackage{ulem}
\usepackage{babel}

\begin{document}

\title{Distribution of high-dimensional orbital angular momentum entanglement
at telecom wavelength over 1km vortex fiber}

\author{Huan Cao$^{1,2}$}

\author{She-Cheng Gao$^{3}$}

\author{Chao Zhang$^{1,2}$}

\author{Jian Wang$^{1,2}$}

\author{De-Yong He$^{1,2}$}

\author{Bi-Heng Liu$^{1,2}$}

\author{Zheng-Wei Zhou$^{1,2}$}

\author{Yu-Jie Chen$^{4}$}

\author{Zhao-Hui Li$^{4}$}
\email{lzhh88@mail.sysu.edu.cn}

\author{Si-Yuan Yu$^{4}$}
\email{yusy@mail.sysu.edu.cn}

\author{Jacquiline Romero$^{5}$}

\author{Yun-Feng Huang$^{1,2}$}
\email{hyf@ustc.edu.cn}

\author{Chuan-Feng Li$^{1,2}$}
\email{cfli@ustc.edu.cn}

\author{Guang-Can Guo$^{1,2}$}

\address{$^{1}$CAS Key Laboratory of Quantum Information, University of Science
and Technology of China, Hefei, 230026, China ~\\
$^{2}$CAS Center For Excellence in Quantum Information and Quantum
Physics, Hefei, 230026, China~\\
$^{3}$Department of Electronic Engineering, College of Information
Science and Technology, Jinan University, Guangzhou, 510632, China~\\
$^{4}$State Key Laboratory of Optoelectronic Materials and Technologies
and School of Electronics and Information Technology, Sun Yat-sen
University, Guangzhou, 510275, China~\\
$^{5}$Centre for Engineered Quantum Systems, School of Mathematics
and Physics, University of Queensland, Queensland 4072, Australia}
\begin{abstract}
 \textcolor{black}{High-dimensional entanglement has demonstrated
potential for increasing channel capacity and resistance to
noise in quantum information processing. However, its distribution
is a challenging task, imposing a severe restriction on its
application. Here we report the first distribution of three-dimensional
orbital angular momentum (OAM) entanglement via a 1-km-long optical
fibre. Using an actively-stabilising phase precompensation technique,
we successfully transport one photon of a three-dimensional OAM entangled
photon pair through the fibre. The distributed OAM entangled state
still shows a fidelity up to 71\% with respect to the three-dimensional
maximal-entangled-state (MES). In addition, we certify that the high-dimensional
quantum entanglement survives the transportation by violating a generalised
Bell inequality, obtaining a violation of $\sim3$ standard deviations
with $I_{3}=2.12\pm0.04$. The method we developed can be extended
to higher OAM dimension and larger distances in principle. Our results
make a significant step towards future OAM-based high-dimensional
long distance quantum communication.}
\end{abstract}
\maketitle

\section{INTRODUCTION}

Increasing the channel capacity and tolerance to noise in quantum communications are strong practical motivations for encoding quantum information in multilevel systems, qudits as opposed to qubits \cite{fujiwara2003exceeding,bechmann2000quantum,gisin2002quantum,aolita2007quantum,ali2007large,d2012complete,nunn2013large,mower2013high,graham2015superdense, Langford2004}. From a foundational perspective, entanglement in higher dimensions exhibits more complex structures \cite{dada2011experimental,romero2012increasing,giovannini2013characterization,krenn2014generation,malik2016multi},
and stronger nonclassical correlations \cite{collins2002bell}. Despite these benefits, the distribution of high-dimensional entanglement is relatively new and remains challenging. We improve on previous works \cite{loffler2011fiber, kang2012measurement} by (1) distributing three-dimensional---rather than two-dimensional---entanglement, and (2) increasing the length of the fibre by three orders of magnitude---achieving the first OAM entanglement distribution via fibre in the kilometer regime.

Inside the laboratory three-level qutrit quantum teleportation    has been achieved \cite{luo2019quantum,hu2019experimental}.  To bring the benefit of qudits into practice, long-distance shared entanglement is required. To this end, a 10-dimensional entangled state has been transmitted over a 24-km long fibre using frequency modes \cite{kues2017chip}; a 4-dimensional entangled state has been transmitted over 100-km long fibre using time-bins \cite{ikuta2018four}.  Hybrid polarisation-time bin entanglement has also been distributed across a 1.2-km free-space intracity link \cite{steinlechner2017distribution}. Distributing entanglement using spatial degrees of freedom, like path and transverse spatial modes, has also been achieved albeit on very modest scales: 4-dimensional path entanglement over a 30-cm long multi-core fibre \cite{lee2017experimental}, and 2-dimensional transverse spatial mode entanglement over a 30-cm hollow photonic crystal fibre \cite{loffler2011fiber} and 40-cm step-index fibre \cite{kang2012measurement}.  Our work improves on these last two results by showing qutrit entanglement transport through fibre in the kilometer regime.

Spontaneous parametric down-conversion (SPDC) naturally conserves both energy and momentum. As a consequence, entanglement in temporal and spatial degrees of freedom (DOF) comes naturally. These two DOFs are both suitable to be used as qudits. We focus on the transverse spatial modes associated with photonic OAM. An OAM state is denoted by $|\ell\rangle$, where $\ell$ is an integer that describes the azimuthal phase dependence corresponding to $\ell\hbar$ of orbital angular momentum \cite{o2002intrinsic}. High-dimensional OAM entanglement upto 50 dimensions can be generated by SPDC  by tuning phase-matching \cite{romero2012increasing}, and even higher OAM states---up to $\ell{=}10,010$---can be entangled with polarisation \cite{fickler2016quantum}. As such, working with OAM or transverse spatial modes, is a viable way of scaling up the dimensionality of entanglement.  Our improving capabilities in preparing \cite{wang2017generation}, measuring \cite{kong2017complete} and processing \cite{babazadeh2017high} these entangled states, together with extensions to multiple parties \cite{erhard2017experimental,malik2016multi} make a strong case for transverse spatial mode as a platform for quantum computation and communication. However, these efforts will be in vain if we are not able to distribute entanglement over long distances.

The sensitivity of OAM to atmospheric turbulence makes it challenging---though not impossible with hybrid entanglement \cite{krenn2015twisted}---to distribute OAM entanglement over free-space. A free-space quantum channel is also subject to weather, line-of-sight, or time of day. Instead, we set out to use an optical fibre to distribute entanglement. Previous works were limited to short fibres of lengths $30{-}40$ cm  and only restricted to two dimensions. \cite{loffler2011fiber, kang2012measurement}.

The signal from a single photon is extremely weak, and hence vulnerable to intermodal mixing or crosstalk. There is a tradeoff between intermodal mixing---the coupling between degenerate modes---and intermodal dispersion---the difference in group velocities for modes of different orders. The latter is a significant factor contributing to decoherence of superposition states even for very short fibres.  Ideally, one could minimise dispersion by using modes from the same order, but doing so increases the intermodal mixing. In \cite{Oxenlowe2019} single photons from an attenuated laser, encoded with OAM and OAM mode superpositions were transmitted through custom-made fibre designed to have low crosstalk between higher-order OAM modes.  We extend their results  by using an entangled photon source. We choose to work with a step-index fibre as in \cite{Carpenter2013} because of their ubiquity in optical communications---we show that a step-index fibre can be used to distribute OAM entanglement over long distances.

We first needed a source of high-dimensional maximally entangled states (MES) for photons at the telecom wavelength. Spontaneous parametric down-conversion (SPDC) is a natural source of photons entangled in their OAM, albeit the entanglement is often non-maximal \cite{torres2003quantum, Langford2004}. The quality of the MES increases with the spiral bandwidth---the number of entangled OAM modes generated. The spiral bandwidth is wider for thin crystals \cite{miatto2012bounds}, but the lower  photon counts (compared to longer crystals) and wider spectral bandwidth present a challenge. Instead of a thin bulk crystal for SPDC, we used a 10-mm-long periodically poled potassium titanyl phosphate (PPKTP) crystal.  We experimentally determined the intermodal dispersion and employed an actively-stabilising precompensation module to eliminate it. With these measures, we successfully transmitted one of the photon pairs through 1 km of fibre and demonstrated the entanglement via generalized bell inequality \cite{collins2002bell} violation of three standard deviations.

\section{EXPERIMENT}

\begin{figure*}
\textcolor{black}{\includegraphics[scale=0.12]{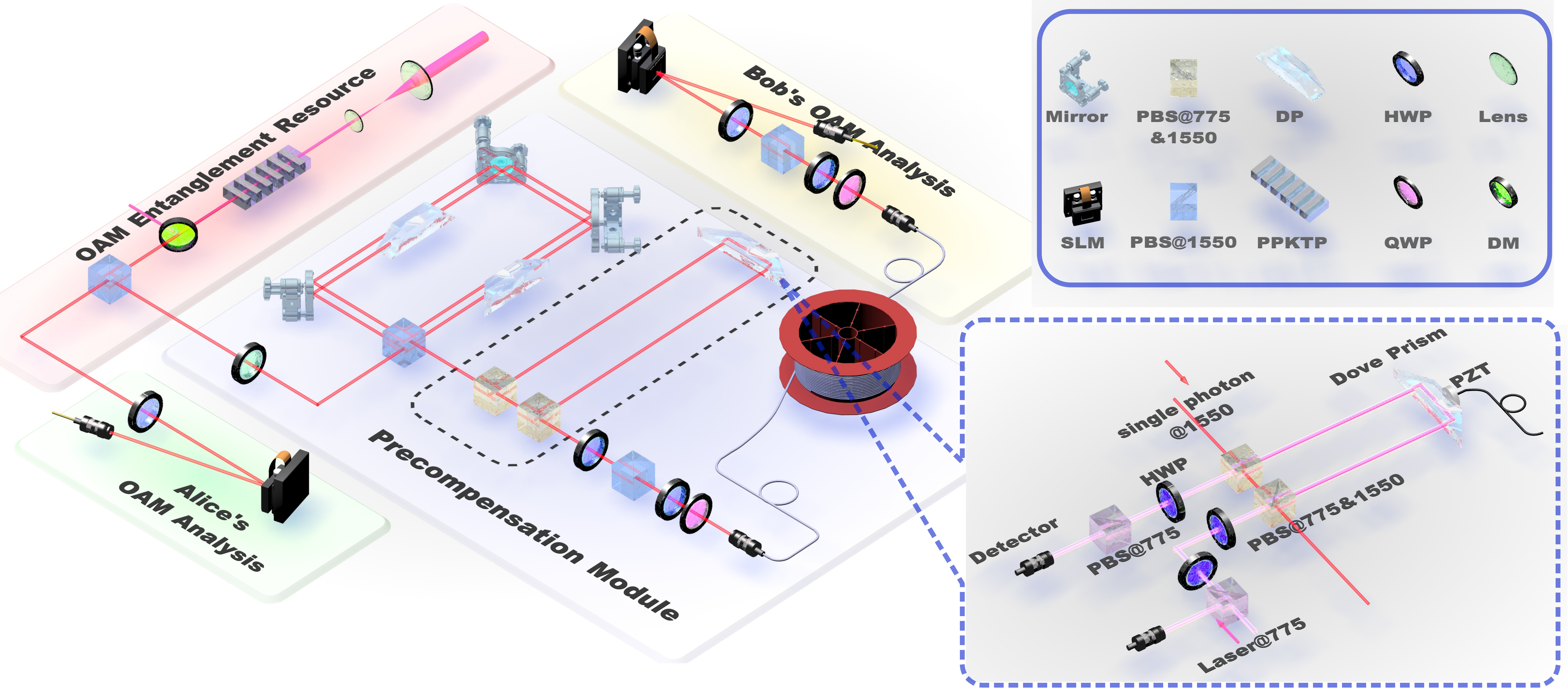}}
\textcolor{black}{\caption{\label{fig:experimental setup}Schematic of the experimental setup. The entangled photons are generated by Type-II SPDC in PPKTP crystal. The idler photon is directly measured (Alice) while the signal photon is fed to a precompensation module and coupled into a 1-kilometer-long OAM fiber and finally analyzed (Bob). The single-photon detectors we use are InGaAs detectors. Here the black dotted box is active locked unbalanced M-Z inteferometer of which the more detailed setup figure is present by the inset. The M-Z inteferometer are both applied to the single photon and classical 775nm reference  beam.} }
\end{figure*}

\textcolor{black}{We now give more details of the experiment: The entangled photons are produced
by degenerate type-II collinear SPDC. Fig. \ref{fig:experimental setup} shows the schematic of the experiment.
A continuous-wave (CW) 775-nm laser beam is coupled to a single mode
fiber (SMF) to obtain a pure fundamental Gaussian mode pump beam. At the output of the SMF,
the pump beam is demagnified by a pair of lenses (not shown in Fig.
\ref{fig:experimental setup}) and is incident on a 10-mm-long periodically
PPKTP crystal. The power of the
pump beam is 36 mw and beam waist on the crystal is $\omega_{0}\backsim175\,\mu m$.
The centre wavelengths of the generated signal and idler photons are $\lambda_{A}{=}\lambda_{B}{=}1550\,nm$, where subscripts
A and B refer to Alice and Bob respectively. The
pump beam is then blocked by a dichroic mirror.}

\begin{figure}
\textcolor{black}{\includegraphics[scale=0.26]{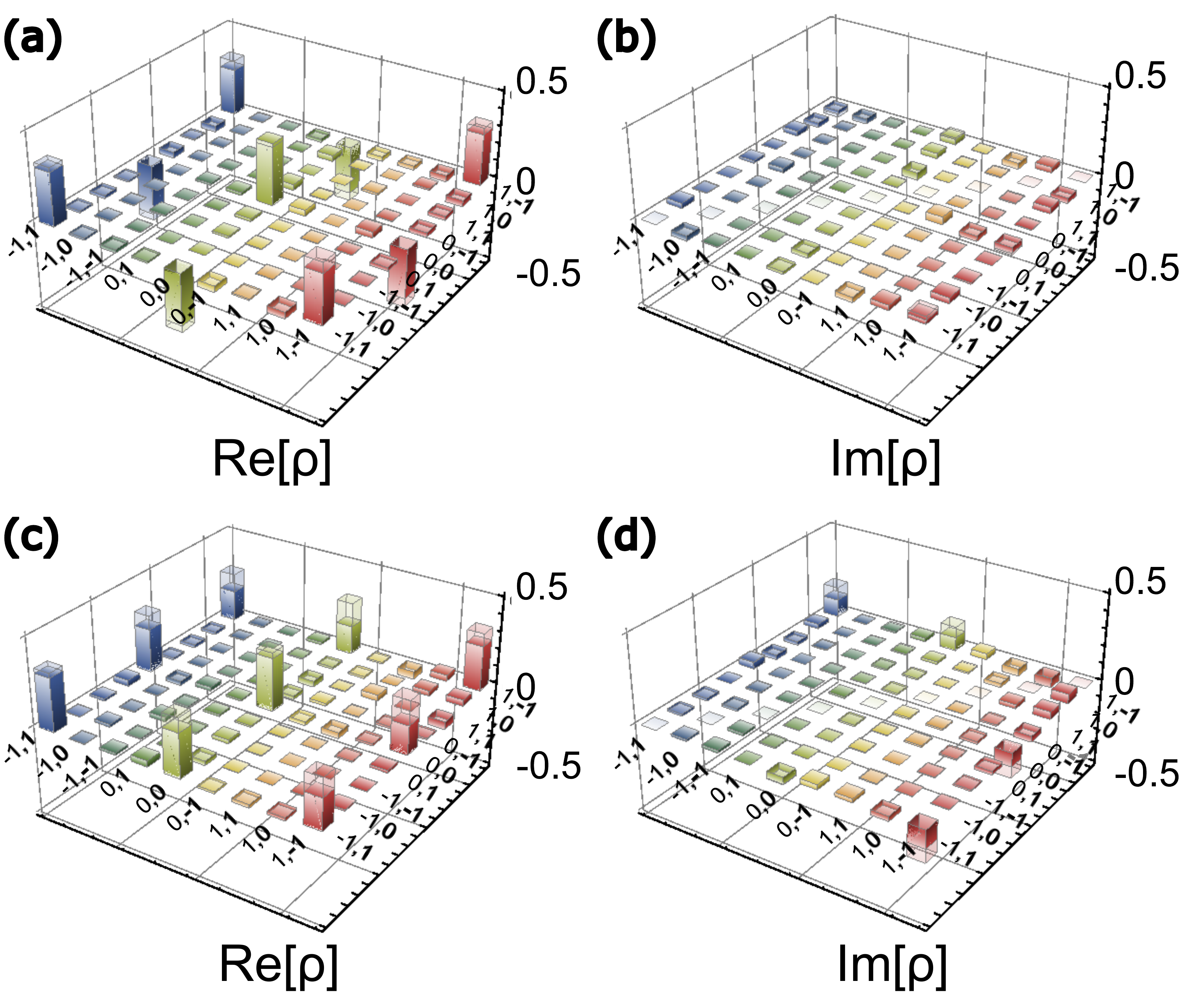}\caption{\label{fig:DM}Reconstructed density matrices before ((a), (b)) and after ((c), (d)) distribution. The left column depicts the real part of reconstructed density matrix while the right part depicts the imaginary part. For comparison, the experimental tomographic matrix is drawn with opaque pillars while its closet maximal entangled state (MES) is drawn with transparent pillars.}
}
\end{figure}

The OAM state of the photon
pairs can be described in terms of Laguerre-Gaussian (LG) modes
following \[ \left|\Psi_{SPDC}\right\rangle =\sum_{\ell,\,p_{s},\,p_{i}}C_{\ell,-\ell}^{p_{s},p_{i}}\left|\ell,\,p_{s}\right\rangle \left|-\ell,\,p_{i}\right\rangle \]
where $C_{\ell,-\ell}^{p_{s},p_{i}}$ denotes the complex weightings of the states $\left|\ell,\,p_{s}\right\rangle $
and $\left|-\ell,\,p_{i}\right\rangle $ the quantum state of signal
and idler respectively \cite{miatto2012spatial}. We limit measurements only to $p_s{=}p_i{=}0$, hence we drop this indices and focus only on the azimuthal mode
indices $\ell$: we analyze in the three-dimensional Hilbert space spanned
by $\ell=0,\pm1$. The post-selected state can be written as
\[
\left|\Psi_{SPDC}\right\rangle =C_{0,0}\left|0\right\rangle \left|0\right\rangle +C_{1,-1}\left|1\right\rangle \left|-1\right\rangle +C_{-1,1}\left|-1\right\rangle \left|1\right\rangle
\]
Here, $C_{0,0},\:C_{1,-1}$ and $C_{-1,1}$ depend on the pump beam
profile, crystal length and ratio between pump waist and down-converted
beam waist \cite{miatto2012bounds,miatto2011full}. The quantum state right after the source was reconstructed via standard quantum state tomography using maximum likelihood estimation \cite{thew2002qudit}, we show the density matrix in
Fig.~\ref{fig:DM} (a), (b).  The fidelity
with respect to three-dimensional MES is $F_{source}{=}0.888\pm0.007$,
and the purity is $P_{source}{=}Tr\left(\rho^{2}\right){=}0.83\pm0.01$.
We note that maximally-entangled states are more easily achieved with thin crystals---these result to a wider spiral spectrum for the same pump characteristics \cite{miatto2012bounds} thus leading to states with higher fidelity to an MES. However, thin crystals also result to lesser photon counts that is detrimental to the signal-to-noise ratio, and wider bandwidth not ideal for distribution.  We thus chose to generate photons in a PPKTP crystal as in \cite{zhou2015quantum} to balance the fidelity and photon counts, and also to give a narrower bandwidth.  Details of the experimental parameters are shown in Supplementary Materials.

The signal and idler photons are separated by a polarising beamsplitter
(PBS). The idler photons are directly analysed by a spatial light
modulator (SLM) via phase-flattening measurements \cite{qassim2014limitations} (Alice in
Fig. \ref{fig:experimental setup}, green region). The signal photon, before going through the 1-km fibre, is
fed to the precompensation module, (red region in Fig. \ref{fig:experimental setup}). This part is critical
in the experiment because the intermodal dispersion among different
OAM modes acts as a dephasing channel, inevitably leading to decoherence.  Therefore, high-dimensional
superpositions and entanglement would be destroyed in a few centimeters \cite{loffler2011fiber,kang2012measurement}.
This is contrast to deterministic transmission of classical OAM information, where one can simply tune the electrical delay to compensate for the delays of the different modes.

We experimentally determine the intermodal
dispersion and devise a setup to reverse it,
to pre-compensate before entering the 1-km fiber (blue region in Fig. \ref{fig:experimental setup}). The precompensation
module consists of two cascaded interferometers and a locking
system. The first interferometer is an OAM sorter which serves as
a parity check to convert the different OAM to polarisation according
to their topological charge $\ell$. We redesigned the OAM sorter from the Mach-Zehnder configuration \cite{zhang2014mimicking} into a Sagnac interferometer
for more robust phase stability. A half wave plate (HWP) is used
to rotate the polarisation of signal photon into $\left(\left|H\right\rangle +\left|V\right\rangle \right)/\sqrt{2}$ before entering the OAM sorter. This first interferometer is designed such that OAM modes with odd topological
charge ($\ell{=}\pm1$) are converted to horizontal polarization, and the
even ones $\left(\ell=0\right)$ are converted to vertical. The second interferometer is an unbalanced Mach-Zehnder (MZ) interferometer designed to separate the different OAM modes into unequal path lengths: the odd OAM modes
enter the short arm and even ones enter the long arm. A dove prism in the long arm is mounted on a translation stage for arm-length tuning to compensate for the intermodal dispersion. To stabilise the path difference between the long and short arms, a phase locking system is used (depicted in Fig. \ref{fig:experimental setup} inset). This is composed of a 775 nm laser beam, two photodetectors, and a PZT (piezoelectric transducer) mounted dove prism. Here the phase between the two arms is locked by a 775 nm classical light separated from the pump laser beam. The power of reference light fed into U-shape interferometer is monitored by a tunable beam splitter consisting of a HWP, a PBS at 775 nm and a detector. Both the single photons and the reference 775 nm beam go through the unbalanced MZ inteferometer, the PBSs at the input and output works for both 775 nm and 1550 nm. A HWP ($22.5^{\circ}$) and PBS combined with the detector at the output of laser beam acquire the feedback power signal. Any fluctuation in path difference perturbs the relative phase between $\left|0\right\rangle$   and $\left|\pm1\right\rangle$ which could be tracked through the readout of the detector. The feedback power signal drives the PZT to stabilize the phase relation with the help of the proportional-integral-derivative (PID) control module. After the cascaded interferometer the single photon is sent to a half-waveplate (HWP) and a polarizing beamspliter
(PBS) to erase any polarisation information.

\begin{figure}
\textcolor{black}{\includegraphics[scale=0.28]{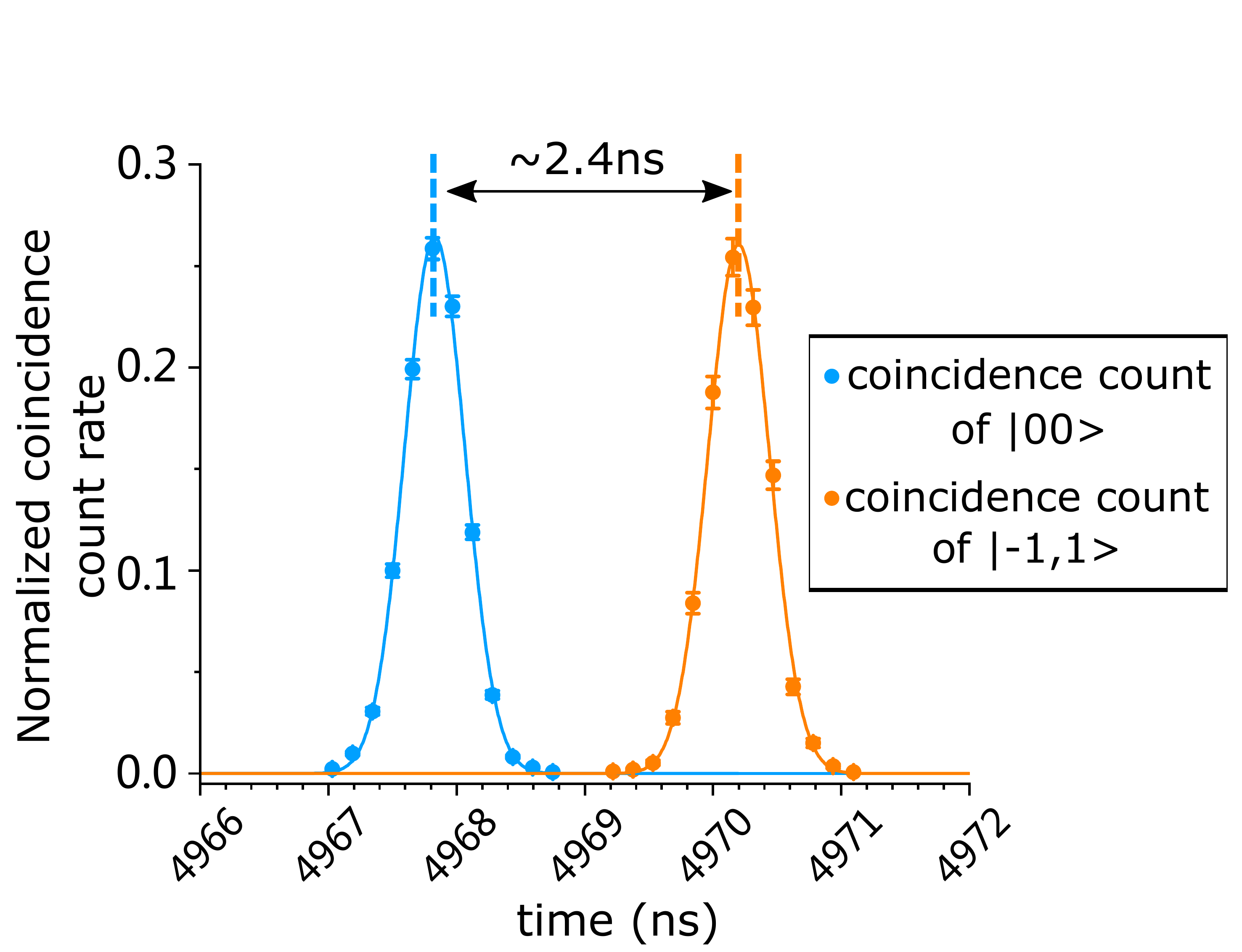}\caption{\label{fig:intermodal dispersion}Intermodal dispersion. The normalized coincidence rate is the ratio between each measured coincidence rate and
their sum. The horizontal coordinate represents the optical delay
between the signal and idler photons. The blue dots represent the
coincidence rates for of $\left|0,0\right\rangle $ and the orange
dots are for $\left|-1,1\right\rangle $. The intermodal dispersion
is estimated by taking the difference in time of the peaks of the Gaussian fits (solid line).}
}
\end{figure}
After precompensation, the signal photon is coupled into
a 1 km long home-designed step-index fibre which supports OAM modes with $\ell=0,\,\pm1,\,\pm2$ (see supplementary material
for details on the fibre). The signal photon is then sent to an SLM for analysis (Bob in Fig.\ref{fig:experimental setup}, yellow region). Because the intermodal dispersion varies as the temperature fluctuates,
we put the step-index fibre in
a sealed adiabatic box with the temperature fluctuation controlled to
within $\pm0.01$ K. Finally, the signal photons are detected over a 0.5 nm bandwidth in Bob's analysis setup.

We obtain the intermodal dispersion by tuning the delay time of the
coincidence module and observing the interval between two coincidence
count events ($\left|0,0\right\rangle $ and $\left|-1,1\right\rangle $), found to be 2.4 ns (See Fig. \ref{fig:intermodal dispersion}
). The accuracy of this is subject only to the temporal resolution of coincidence module (156 ps),
time jitter only broadens the shape of the coincidence curves but does not alter the interval
of their peaks.
\section{CHARACTERIZATION AND CERTIFICATION OF HIGH-DIMENSION NONLOCALITY}

The reconstructed density matrix of the entangled
state after the signal photon goes through the fibre is presented in Fig. \ref{fig:DM}(c), (d). The 81 measurement settings required for quantum tomography are represented
by projectors $p_{i}\otimes p_{j}(i=1,2,\ldots,9)$, where $p_{k}=\left|\varphi_{k}\right\rangle \left\langle \varphi_{k}\right|$.
$\left|\varphi_{k}\right\rangle $ is selected from the following
set: $\{\left|-1\right\rangle $, $\left|0\right\rangle $, $\left|1\right\rangle $,
$\left(\left|0\right\rangle +\left|-1\right\rangle \right)/\sqrt{2}$,
$\left(\left|0\right\rangle +\left|1\right\rangle \right)/\sqrt{2}$,
$\left(\left|0\right\rangle +i\left|-1\right\rangle \right)/\sqrt{2}$,
$\left(\left|0\right\rangle -i\left|1\right\rangle \right)/\sqrt{2}$,
$\left(\left|-1\right\rangle +\left|1\right\rangle \right)/\sqrt{2}$,
$\left(\left|-1\right\rangle +i\left|1\right\rangle \right)/\sqrt{2}\}$\cite{thew2002qudit}.

To confirm the quality of the distributed entanglement, we evaluate the
fidelity with respect to the three-dimensional MES. The MES can be represented
by
\[
\left|\Psi_{MES}\left(\theta,\varphi\right)\right\rangle =\left(e^{i\theta}\left|-1\right\rangle \left|1\right\rangle +\left|0\right\rangle \left|0\right\rangle +e^{i\varphi}\left|1\right\rangle \left|-1\right\rangle \right)/\sqrt{3}
\]
with different values of $\theta$ and $\varphi$ corresponding to different MESs. We performed a global search for the MES
that would give the highest fidelity. With $\theta{=}0.98\times2\pi$ and $\varphi{=}0.07\times2\pi$ we obtain a fidelity $F=\left\langle \Psi_{MES}\left(\theta,\varphi\right)\right|\rho\left|\Psi_{MES}\left(\theta,\varphi\right)\right\rangle =0.71\pm0.02$,
and purity $P=Tr\left(\rho^{2}\right)=0.56\pm0.03$.

This fidelity suffices to certify entanglement beyond two dimension \cite{fickler2014interface}, since the overlap between the obtained state with respect to ideal three-dimensional MES cannot be achieved by an entangled state with lower than three dimensions (see Supplementary Material).
To further illustrate three-dimensional entanglement we violated a generalized Bell inequality, the Collins-Gisin-Linden-Massar-Popescu (CGLMP) inequality, for qutrits.

The Bell expression in three dimension is \cite{collins2002bell}
\begin{align*}
I_{3}\equiv & +[P\left(A_{1}=B_{1}\right)+P\left(B_{1}=A_{2}+1\right)\\
 & \quad+P\left(A_{2}=B_{2}\right)+P\left(B_{2}=A_{1}\right)]\\
 & -[P\left(A_{1}=B_{1}-1\right)+P\left(B_{1}=A_{2}\right)\\
 & \quad+P\left(A_{2}=B_{2}-1\right)+P\left(B_{2}=A_{1}-1\right)],
\end{align*}
where
\[
P\left(A_{a}=B_{b}+k\right)\equiv\sum_{j=0}^{2}P\left(A_{a}=j,\,B_{b}=j+k\,mod2\right)
\]
We searched for the specific measurement settings $A_{1},$ $A_{2}$
and $B_{1}$, $B_{2}$ to maximally violate the inequality, $I_{3}\leq2$, imposed by local realism. We obtained $I_{3}=2.12\pm0.04$.
The violation exceeds the classical bound by about 3 standard deviations.
Bell-type inequalities such as the CGLMP inequality we use can be considered as entanglement witnesses \cite{hyllus2005relations}, thus the violation that we show sufficiently proves the existence of three-dimensional entanglement.

\section{DISCUSSION AND CONCLUSION}

In summary, we have distributed OAM entanglement over 1 km of fibre, three orders of magnutide over previous work \cite{loffler2011fiber,kang2012measurement}.  The challenging task of maintaining a stable phase relation in the entangled state when one photon undergoes evolution in an optical fibre was overcome by: (1) keeping the temperature of the fibre stable to avoid temperature-induced phase fluctuations in the fibre, and (2) precompensating for the intermodal dispersion. For the latter, we redesigned an OAM sorter \cite{zhang2014mimicking} into a Sagnac cofiguration for better phase stability, and used an unbalanced Mach-Zender interferometer to introduce different delays  between the $|0\rangle$ and $|{\pm{1}}\rangle$ states. The dispersion between $|1\rangle$ and $|{-1}\rangle$ is extremely small even for 1-km-length fiber  and made even more negligible by narrowing the bandwidth of the measurement such that this dispersion becomes negligible compared to the long coherence time. With our measures, we are able to certify three-dimensional entanglement via a fidelity to the the MES of 0.71, and a violation of a CGLMP inequality.

\subsection*{\textcolor{black}{}}
\begin{acknowledgments}
We thank Zongquan Zhou , Zhiyuan Zhou, Xiao Liu, Zhaodi Liu for beneficial discussion for experiment, and Andrew White for careful reading of the manuscript.
This work was supported by the National Natural Science Foundation of China (Grant Numbers 61490711, 61327901, 11734015, 11804330, 11874345, 11821404, 11774335, 11704371), the National Key Research and Development Program of China (Grants Number 2017YFA0304100), Anhui Initiative in Quantum Information Technologies (AHY070000), Local Innovative and Research Teams Project of Guangdong Pearl River Talents Program
(2017BT01X121), Key Research Program of Frontier Sciences, CAS (No.
QYZDY-SSW-SLH003), the National Youth Top Talent Support Program of
National High level Personnel of Special Support Program, the Fundamental
Research Funds for the Central Universities (Grants Number WK2030020019). JR is supported by a Westpac Research Fellowship and Australian Research Council Centre of Excellence for Engineered Quantum Systems (EQUS, CE170100009).
\end{acknowledgments}

\clearpage
\section*{Supplementary Material for : Distribution of high-dimensional orbital angular momentum entanglement at telecom wavelength over 1km of optical fibre}

\subsection{Source of OAM-entangled Photons}

The orbital angular momentum (OAM) is conserved in the spontaneous
parameter down-conversion process (SPDC) and hence SPDC is the most
convenient and efficient method to generate high-dimensional OAM entanglement. However, in practice, such a method generates non-maximally
entangled states. The fidelity of the generated state to the maximally
entangled state (MES) is related to the crystal length $L$, pump
waist $\omega_{p}$, signal (idle) photon waist $\omega_{s}$ ($\omega_{i}$)
\citep{miatto2012bounds}. Generally, thin crystals such
as beta barium borate ($\beta-$BBO) lead to states that are closer to the MED owing to the wider quantum spiral spectrum \citep{torres2003quantum}.

In our experiment, we are limited by our InGaAs detectors, which for 1550 nm have about 12\% efficiency. Instead of using a thin bulk crystal like BBO, we used a 10-mm-long periodically poled potassium titanyl
phosphate (PPKTP) crystal, which gives us more photon counts and also a narrower spectral bandwidth.   The pump beam is demagnified and collimated by a pair
of lenses with the focal length $f=125\,mm$ and $f=25.4\,mm$, before it impinges on the PPKTP crystal with a beam waist
of $\omega_{p}\sim175\,\mu m$. After that PPKTP is
directly imaged to two SLMs for analysis using a lens with the focal
length $f=150\,mm$.

To improve the accuracy, the hologram embedded in the SLM contains intensity and
phase modulations\citep{bolduc2013exact}. The SLM
hologram is given by

\begin{equation}
\Psi\left(\rho,\phi\right)=\mathcal{L}\left(\rho,\phi\right)Mod\left(\mathcal{F}\left(\rho,\phi\right)+2\pi x/\Lambda,2\pi\right)\label{eq:hologram},
\end{equation}
with the two modulation functions described by,

\begin{equation}
\begin{array}{c}
\mathcal{L}\left(\rho,\phi\right)=1+\frac{1}{\pi}sinc^{-1}\left(A\right)\\
\mathcal{F}\left(\rho,\phi\right)=\phi\pi\mathcal{L}.
\end{array}\label{eq:modified function}
\end{equation}
Here, $\phi$ and $A$ are the phase and amplitude distribution of
desired OAM state respectively; $2\pi x/\Lambda$ denotes the phase
of the blazed grating with period of $\Lambda$ along the x coordinates.
Here $\mathcal{L}$ is a normalized bounded positive function of amplitude
, i.e, $0\leq\mathcal{L}\leq1$, which corresponds to the mapping
of the phase depth to the diffraction efficiencies of the spatially
dependent blazing function.

The single mode fiber (SMF) accepts only the fundamental Gaussian mode, and is connected to a single photon detector
(SPD). Because the conversion to the fundamental Gaussian mode is not perfect---some intensity remains outside central bright spot of the fundamental mode \cite{qassim2014limitations}---
there is a difference of detecting efficiencies among different OAM states. We carefully adjusted the pair of lenses ($f=125\,mm$) to minimise the differences.

In order to characterize the quality of the entangled source, we
calculate the fidelity with respect to a 3*3 maximally entangled state (MES).
We performed a global search to find one MES which
is closest to the reconstructed quantum state. The MES is described
by
\[
\left|\varPsi_{MES}\left(\theta,\varphi\right)\right\rangle =e^{i\theta}\left|-1\right\rangle \left|1\right\rangle +\left|0\right\rangle \left|0\right\rangle +e^{i\varphi}\left|1\right\rangle \left|-1\right\rangle
\]
 With $\theta=1.02\pi$, $\varphi=0.98\pi$,
the highest fidelity between the constructed density matrix $\rho$
and $\varPsi_{MES}$ is $\left\langle \varPsi_{MES}\right|\rho\left|\varPsi_{MES}\right\rangle =0.888\pm0.007$

\subsection{Eigenmodes in Fiber}

The OAM modes are not eigenmodes of the step-index fiber, but it is possible to decompose OAM modes into a superposition of eigenmodes in our step-index fiber. For example, we write the $\left|\ell\right|=1$ mode as following
\[
\begin{cases}
\mathrm{OAM_{\pm1}^{\pm}} & =\mathrm{HE_{2,1}^{even}}\pm i\mathrm{HE_{2,1}^{odd}}\\
\mathrm{OAM_{\pm1}^{\mp}} & =\mathrm{TM_{0,1}}\pm i\mathrm{TE_{0,1}}
\end{cases}
\]
The superscripts $\pm$ over the OAM represent the right(+) or left(-)
circular polarization of photonic state. The four vector modes $\mathrm{HE_{2,1}^{even}}$,
$\mathrm{HE_{2,1}^{odd}}$, $\mathrm{TM_{0,1}}$ and $\mathrm{TE_{0,1}}$
, designated as $\mathrm{LP_{11}}$ modes in the scalar approximations,
are first higher order eigenmodes of which the effective refractive
index are nearly degenerate. The two $\mathrm{HE_{21}}$ modes are
strictly degenerate and have an effective refractive index $n_{eff}$
distinct from $\mathrm{TE_{01}}$ and $\mathrm{TM_{01}}$. The effective
index of $\mathrm{TE_{01}}$ and $\mathrm{TM_{01}}$ are also slightly
different. Nevertheless, discrepancy of $n_{eff}$ among these four
modes is extremely small.

In our experiment, the four vector modes are all involved. The near-
degeneracy of the modes other than the desired modes makes crosstalk likely. Decoherence between
$\ell=1$ and $\ell=-1$ is small because of the small intermodal dispersion.
In the main text, the precompensation module is used to compensate for
intermodal dispersion (2.4 ns) between $\ell=0$ and $\ell=\pm1$.
As for dispersion between $\ell=1$ and $\ell=-1$, we
employed a wave-division-multiplexing (WDM) with a 0.5-nm channel bandwidth,
making coherent time much larger such that our measurements are insensitive to the dephasing caused by this intermodal dispersion.

Our fibre is a step-index fiber. It has a core radius of ${\sim}9.5\,\mu m$, a cladding radius of ${\sim}62.5\,\mu m$, and a core cladding refractive index difference of ${\sim}5\times10^{-3}$, same fibre that we used in our previous work \citep{wu2017all,xie2018integrated}. This fibre supports four modes groups $\left(\mathrm{LP_{01},\,LP_{11},\,LP_{21},\,LP_{02}}\right)$, but in our work only two modes groups, i.e. $\mathrm{LP_{01}}$ and $\mathrm{LP_{11}}$ , are used. The effective refractive index
of the each LP mode $\left(\mathrm{LP_{01},\,LP_{11},\,LP_{21},\,LP_{02}}\right)$,
corresponding to the vector core mode $(\mathrm{HE_{11},\,TE_{01},\,HE_{21},\,TM_{01},\,EH_{11},\,HE_{31},\,HE_{12}})$,
is calculated using a full-vector finite-element mode solver, and
is given in \citep{wu2017all}.

\subsection{Bipartite Witness for D-dimensional Entanglement.}

 \begin{table*}[htbp]
\centering
\caption{\bf tomographic measurement result of distributed state after vortex fiber. The reconsctructed density matrix is presented in main text.}
\begin{tabular}{c|ccccccccc}
 &  &  &  &  &  &  &  &  & \tabularnewline
 & $\left|-1\right\rangle $ & $\left|0\right\rangle $ & $\left|1\right\rangle $ & $\frac{\left|0\right\rangle +\left|-1\right\rangle }{\sqrt{2}}$ & $\frac{\left|0\right\rangle +\left|1\right\rangle }{\sqrt{2}}$ & $\frac{\left|0\right\rangle +i\left|-1\right\rangle }{\sqrt{2}}$ & $\frac{\left|0\right\rangle -i\left|-1\right\rangle }{\sqrt{2}}$ & $\frac{\left|1\right\rangle +\left|-1\right\rangle }{\sqrt{2}}$ & $\frac{\left|1\right\rangle -i\left|-1\right\rangle }{\sqrt{2}}$\tabularnewline
 &  &  &  &  &  &  &  &  & \tabularnewline
\hline
\hline
 &  &  &  &  &  &  &  &  & \tabularnewline
$\left|-1\right\rangle $ & 196 & 9 & 17 & 121 & 8 & 132 & 27 & 137 & 101\tabularnewline
 &  &  &  &  &  &  &  &  & \tabularnewline
\hline
\hline
 &  &  &  &  &  &  &  &  & \tabularnewline
$\left|0\right\rangle $ & 8 & 261 & 9 & 119 & 85 & 66 & 106 & 10 & 5\tabularnewline
 &  &  &  &  &  &  &  &  & \tabularnewline
\hline
\hline
 &  &  &  &  &  &  &  &  & \tabularnewline
$\left|1\right\rangle $ & 24 & 14 & 182 & 26 & 146 & 31 & 146 & 77 & 76\tabularnewline
 &  &  &  &  &  &  &  &  & \tabularnewline
\hline
\hline
 &  &  &  &  &  &  &  &  & \tabularnewline
$\frac{\left|0\right\rangle +\left|-1\right\rangle }{\sqrt{2}}$ & 148 & 85 & 9 & 229 & 44 & 122 & 59 & 72 & 49\tabularnewline
 &  &  &  &  &  &  &  &  & \tabularnewline
\hline
\hline
 &  &  &  &  &  &  &  &  & \tabularnewline
$\frac{\left|0\right\rangle +\left|1\right\rangle }{\sqrt{2}}$ & 16 & 115 & 92 & 23 & 162 & 69 & 98 & 49 & 44\tabularnewline
 &  &  &  &  &  &  &  &  & \tabularnewline
\hline
\hline
 &  &  &  &  &  &  &  &  & \tabularnewline
$\frac{\left|0\right\rangle +i\left|-1\right\rangle }{\sqrt{2}}$ & 101 & 115 & 21 & 84 & 61 & 26 & 96 & 86 & 72\tabularnewline
 &  &  &  &  &  &  &  &  & \tabularnewline
\hline
\hline
 &  &  &  &  &  &  &  &  & \tabularnewline
$\frac{\left|0\right\rangle -i\left|-1\right\rangle }{\sqrt{2}}$ & 19 & 116 & 72 & 79 & 56 & 82 & 35 & 21 & 42\tabularnewline
 &  &  &  &  &  &  &  &  & \tabularnewline
\hline
\hline
 &  &  &  &  &  &  &  &  & \tabularnewline
$\frac{\left|1\right\rangle +\left|-1\right\rangle }{\sqrt{2}}$ & 127 & 16 & 83 & 51 & 82 & 63 & 105 & 147 & 106\tabularnewline
 &  &  &  &  &  &  &  &  & \tabularnewline
\hline
\hline
 &  &  &  &  &  &  &  &  & \tabularnewline
$\frac{\left|1\right\rangle -i\left|-1\right\rangle }{\sqrt{2}}$ & 97 & 13 & 86 & 74 & 84 & 76 & 107 & 141 & 23\tabularnewline
 &  &  &  &  &  &  &  &  & \tabularnewline

\hline
\end{tabular}
  \label{tab:tomographyresult}
\end{table*}

The 81 measurement settings required for quantum tomography are represented by projectors $p_{i}\otimes p_{j}(i=1,2,\ldots,9)$, where $p_{k}=\left|\varphi_{k}\right\rangle \left\langle \varphi_{k}\right|$.
The set of $\left|\varphi_{k}\right\rangle $ is given in the main text. The measurement results are presented in Table\ \ref{tab:tomographyresult}. We performed quantum state tomography and obtained the fidelity
$F$ both before and after distribution. We employed the method developed in \citep{Fickler2014} to certify the 3-dimensional entanglement.
In the following text, we bound the maximal overlap between the chosen
high-dimensional state and states with a bounded Schmidt rank $d$.
If the fidelity reveals a higher overlap than this bound, the justification
of at least $\left(d+1\right)$-dimensional entanglement is proven.

The Schimidt decomposition of the assumed high-dimensional state is
described as $\left|\varphi\right\rangle =\sum_{i=1}^{D}\lambda_{i}\left|ii\right\rangle $
with the coefficients in a decreasing order $\left|\lambda_{1}\right|\geq\left|\lambda_{2}\right|\geq\cdots\geq\left|\lambda_{D}\right|$,
where $i$ denotes the different OAM state and $D$ the dimension
of the Hilbert space. The witness for $d$-dimensional entanglement
is constructed by comparing the two fidelities
\begin{eqnarray*}
F & = & Tr\left(\rho\left|\varphi\right\rangle \left\langle \varphi\right|\right)\\
F_{d} & = & \max_{\phi_{d}}\left|\left\langle \phi_{d}|\varphi\right\rangle \right|^{2}
\end{eqnarray*}
where $\rho$ is the density matrix after distribution and $\left|\phi_{d}\right\rangle =\sum_{m,n=1}^{D}\alpha_{mn}\left|mn\right\rangle $
represents states with a bounded Schmidt rank $d$. The global search
for maximizing the $F_{d}$ implies that $F\geq F_{d}$ could not
be satisfied by a $d$-dimensional entangled state. In other word,
the generated bipartite system is at least $\left(d+1\right)$-dimensionally
entangled.

We directly calculate $F_{d}$ as the maximal overlap:
\begin{eqnarray*}
F_{d} & = & \max_{\phi_{d}}\left|\left(\left\langle mn\right|\sum_{m,n=1}^{D}\alpha_{mn}^{\ast}\right)\left(\sum_{i=1}^{D}\lambda_{i}\left|ii\right\rangle \right)\right|^{2}\\
 & = & \max_{\phi_{d}}\left|Tr\left(\sum_{i,m,n=1}^{D}\left\langle m|i\right\rangle \left\langle n|i\right\rangle \alpha_{mn}^{\ast}\lambda_{i}\right)\right|^{2}
\end{eqnarray*}
We introduce two operators in the form of
\begin{eqnarray*}
U & = & c_{mn}\left|m\right\rangle \left\langle n\right|\\
P_{d}U^{\ast} & = & U^{\ast}
\end{eqnarray*}
where $P_{d}$ is a rank $d$-projector which always exists if $B^{\ast}$
is of rank d, as $\left|\phi_{d}\right\rangle $ is also of Schmidt
rank d. Combining these equations, we have
\begin{eqnarray*}
F_{d} & = & \max_{\phi_{d}}\left|Tr\left(U\dagger\sum_{i=1}^{D}\lambda_{i}\left|i\right\rangle \left\langle i\right|\right)\right|^{2}\\
 & = & \max_{\phi_{d}}\left|Tr\left(P_{d}U\dagger\sum_{i=1}^{D}\lambda_{i}\left|i\right\rangle \left\langle i\right|\right)\right|^{2}\\
 & = & \max_{\phi_{d}}\left|Tr\left(U\dagger\sum_{i=1}^{D}\lambda_{i}\left|i\right\rangle \left\langle i\right|P_{d}\right)\right|^{2}
\end{eqnarray*}
Note that for the inner product $\left\langle A,B\right\rangle \equiv Tr\left(A,B\dagger\right)$,
taking advantage of Cauchy-Schwarz inequality $\left(\left|\left\langle A,B\right\rangle \right|^{2}\leq\left\langle A,A\right\rangle \left\langle B,B\right\rangle \right)$, the
upper bound of $F_{d}$ is found to be
\[
F_{d}\leq\max_{\phi_{d}}Tr\left(BB\dagger\right)Tr\left(P_{d}\sum_{i=1}^{D}\left|\lambda_{i}\right|^{2}\left|i\right\rangle \left\langle i\right|P_{d}\right).
\]
Because $Tr\left(BB\dagger\right)=\sum_{m,n=1}^{D}c_{mn}c_{nm}^{\ast}\leq1$
and choosing $P_{d}=\sum_{i=1}^{d}\left|i\right\rangle \left\langle i\right|$,
we get the upper bound of $F_{d}$ for $d$-dimensional entangled
states with a simple fomula
\[
F_{d}\leq\sum_{i=1}^{d}\left|\lambda_{i}\right|^{2}
\]
By choosing a specific $\left|\phi_{d}\right\rangle =\left(1/\sqrt{\sum_{i=1}^{d}\left|\lambda_{i}\right|^{2}}\right)\sum_{i=1}^{d}\lambda_{i}\left|ii\right\rangle $,
we find that
\[
F_{d}\geq\sum_{i=1}^{d}\left|\lambda_{i}\right|^{2}
\]
Thus we find a tight bound for witness of $\left(d+1\right)$-dimensional
entanglement
\[
F_{d}=\max_{\phi_{d}}\left|\left\langle \phi_{d}|\varphi\right\rangle \right|^{2}=\sum_{i=1}^{d}\left|\lambda_{i}\right|^{2}
\]
For $d=2$, this upper bound is found to be $\frac{2}{3}$. With the fidelity $F=0.71\pm0.02$ that we experimentally obtained, we conclude that our results can only be explained by a state that is at least three-dimensional entangled
state
\[
F_{d}>\sum_{i=1}^{d}\left|\lambda_{i}\right|^{2}
\]
The witness also holds for mixed state,x which would only lower the
bound due to the convexity of fidelity.

\subsection{Quantum Process Tomography}

The entire state-transfer process (precompensation and evolution in
vortex fiber combined) can be represented by a quantum process $\chi$
\citep{o2004quantum}. The output state can be described by
\[
\rho_{out}=\sum_{m,n=1}^{9}\chi_{mn}\lambda_{m}\rho_{in}\lambda_{n}^{\dagger}
\]
where $\rho_{in}$ is the input state and $\lambda_{m}$ is the basis
of qutrit operators. The process matrix $\chi$ can be identified
by measuring the output $\rho_{in}$ for a series of input state $\rho_{in}$.
The input state are chosen from the states set $|\phi_k\rangle$ (see Table\ \ref{tab:tomographyresult}).

The corresponding complete operator basis are presented
as follows \cite{thew2002qudit}:
$
\lambda_{1}{=}\left\{ \begin{array}{ccc}
1 & 0 & 0\\
0 & 1 & 0\\
0 & 0 & 1
\end{array}\right\} ,
\lambda_{2}{=}\left\{ \begin{array}{ccc}
0 & 1 & 0\\
1 & 0 & 0\\
0 & 0 & 0
\end{array}\right\} $,$\lambda_{3}{=}\left\{ \begin{array}{ccc}
0 & -i & 0\\
i & 0 & 0\\
0 & 0 & 0
\end{array}\right\} $,
\\
$\lambda_{4}{=}\left\{ \begin{array}{ccc}
1 & 0 & 0\\
0 & -1 & 0\\
0 & 0 & 0
\end{array}\right\} $,$\lambda_{5}{=}\left\{ \begin{array}{ccc}
0 & 0 & 1\\
0 & 0 & 0\\
1 & 0 & 0
\end{array}\right\} $,$\lambda_{6}{=}\left\{ \begin{array}{ccc}
0 & 0 & -i\\
0 & 0 & 0\\
i & 0 & 0
\end{array}\right\} $,\\
$\lambda_{7}{=}\left\{ \begin{array}{ccc}
0 & 0 & 0\\
0 & 0 & 1\\
0 & 1 & 0
\end{array}\right\} $,$\lambda_{8}{=}\left\{ \begin{array}{ccc}
0 & 0 & 0\\
0 & 0 & -i\\
0 & i & 0
\end{array}\right\} $,\\$\lambda_{9}{=}\left\{ \begin{array}{ccc}
1 & 0 & 0\\
0 & 1 & 0\\
0 & 0 & -2
\end{array}\right\} /\sqrt{3}$.

\begin{figure}
\includegraphics[scale=0.28]{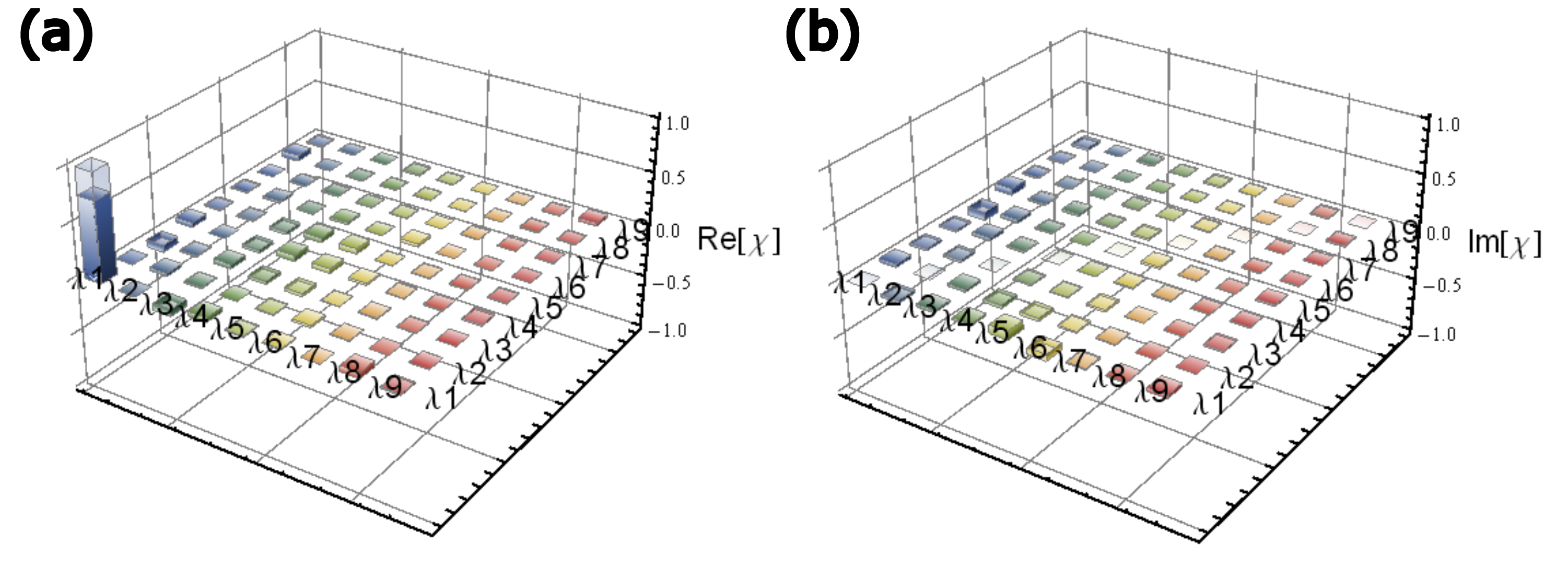}
\caption{\label{fig:process}Process matrix of state transfer process (combination
of precompensation and inside-fiber evolution). (a) and (b) represent
the real and imaginary part of process matrix respectively, with the
opaque histogram denoting the experimental tomographic matrix and
transparent histogram denoting the ideal identity matrix $\chi_{0}=\mathcal{I}$.}
\end{figure}

For an ideal situation, the state transfer process should be an identity
matrix $\chi_{0}=\mathcal{I}$, regardless of the phase shift in fiber
which could be removed by redefining the qutrit basis. Here we applied
the process tomography to the state-transfer process. A wide bandwidth
attenuated laser pass the wave division multiplexing to narrow the
bandwidth to 0.5 nm (as in the main text), the attenuated
single photons are emitted from the coupler in Alice's part, and sent
to the SLM for input state preparation. The PPKTP was replaced by a mirror, and the photons were sent in the reverse direction (from Alices's detector side to mirror). The mirror reflects the photon into the arm of the precompensation module where. The reconstructed process matrix $\chi$ is
shown in Fig.\ref{fig:process}. The fidelity is calculated by $F=\left[\mathrm{Tr}\left(\sqrt{\sqrt{\chi}\chi_{0}\sqrt{\chi}}\right)\right]^{2}=0.747$.
The process is different from identity, which would have been the case ideally. The discrepancy comes from several factors:
According to our experiment data, there still is  12.8\% crosstalk
between the OAM modes. This crosstalk comes from the mode mismatch
in the optical fibre. As is discussed above, in our experiment three OAM
modes ($\ell=0,\pm1$) relate to six eigenstate of $LP_{00}$ and
$LP_{11}$ mode groups in the fiber, crosstalk between degenerate
or near degenerate state is likely. The spatial mode is also sensitive to
any imperfections in the optical surfaces. This affects mode matching into
the fibre. These factors causes are
all experimental imperfections rather than a
drawback of our method in principle.

\bibliographystyle{apsrev}
\bibliography{OAM_3d1km}
\end{document}